\documentclass[a4paper,prb,twocolumn,showpacs,preprintnumbers,amsmath,amssymb]{revtex4}

\usepackage{graphicx}
\usepackage{dcolumn}
\usepackage{bm}
\usepackage{nicefrac} 
\usepackage{amsmath}

\newcommand{\VEV}[1]{\left\langle {#1}\right\rangle} 
\newcommand{\abs}[1]{\left\vert {#1}\right\vert}

\def\rmd{{\rm d}}
\def\rme{{\rm e}}
\def\conv{\mbox{\rm conv}}
\newcommand{\Pad}[1]{{#1} ^{\rm P}  }
\newcommand{\dft}[1]{{\cal F}\left\{{#1}\right\}}
\newcommand{\idft}[1]{{\cal F}^{-1}\!\left\{{#1}\right\}}
\newcommand{\modN}[1]{{\rm mod}(#1,N)}
\newcommand{\moddN}[1]{{\rm mod}(#1,2N)}
\newcommand{\commentout}[1]{}
\def\numparts{{}}
\def\endnumparts{{}}
\newcommand{\eref}[1]{Eq.\ (\ref{#1})}
\newcommand{\neref}[1]{(\ref{#1})}
\def\m{\hbox{$\phantom{-}$}}
\def\0{\hbox{$\phantom{0}$}}

\def\fl{{}}

\def\addrParma{Dipartimento di Fisica e 
Scienze della Terra,
Universit\`a degli Studi di Parma, Viale G.\ Usberti 7A I-43100 Parma, Italy}

\begin{document}

\title[A numerical method for the dynamic muon relaxation function]{A numerical method to calculate the muon relaxation function in the presence of diffusion}

\author{G. Allodi}
\email{Giuseppe.Allodi@fis.unipr.it}
\author{R. De Renzi}
\affiliation{\addrParma}


\begin{abstract}

We present an accurate and efficient method to calculate  the effect of random fluctuations of the local field at the muon, for instance in the case muon diffusion, within the framework of the strong collision approximation.
The method is based on a reformulation of the  Markovian process over a discretized time base, leading to a summation equation for the muon polarization function which is solved by discrete Fourier transform. The latter is 
formally analogous, though not identical, to the integral equation of the original continuous-time  model, solved by Laplace transform.
With real-case parameter values, the solution of the discrete-time strong collision model
is found to approximate the continuous-time solution with excellent accuracy even with a coarse-grained time sampling. 
Its calculation by the fast Fourier transform algorithm is very efficient and suitable for real time fitting of experimental data even on a slow computer.

\end{abstract}


\maketitle
\section{Introduction}
\label{sec:intro}
One of the greatest benefits 
of muon spin rotation ($\mu$SR) as a local probe of magnetism in condensed matter is its capability of detecting randomly distributed static magnetic fields even in the absence of a net bias field. 
This makes $\mu$SR the technique of choice for the study of disordered  magnets or weakly magnetic systems, such as spin and cluster-spin glasses. 
In most cases, purely static magnetic disorder is adequately  accounted for by a classical distribution of random fields, whence the longitudinal muon spin polarization function $G(t)\equiv\VEV{S_z(t)S_z(0)}/\VEV{S_z(0)^2}$ is calculated by averaging the precession waveform of a muon in a random local field over the muon ensemble. For instance a Gaussian distribution, yielding 
the well-known static Kubo-Toyabe function, \cite{KT1967} suitably describes the muon depolarization by the static dipolar fields of nuclei. 

The effect of time-dependent fluctuations on an otherwise static random distribution of fields at the muon can be easily accounted for in the limit of very rapid  fluctuations (the so-called narrowing limit), whereby they produce simple exponential muon spin relaxations. 
The intermediate case between the narrowing limit and purely static 
fields requires however a detailed modelling of the dynamical processes perturbing the instantaneous field at the muon. 
The simplest dynamical model, suitable e.g.\ to describe the effect of muon diffusion,  is based on a {\it strong collision} approximation, \cite{kuboSCM}  dealing with the muon spin evolution in the form of a Markovian process. 
Such a model yields a recursion series for $G(t)$, whose summation leads to an integral equation, which is solved in principle by Laplace transform. \cite{hayano}
However, an exact analytical solution of the strong collision model (SCM) with an arbitrary distribution of static fields cannot be obtained. 
In the most general case,  including that of a Gaussian distribution corresponding to the dynamical Kubo-Toyabe function, Laplace transforms have to be calculated and inverted numerically. 
This makes the Laplace transform method impractical, especially when the free parameters of the model 
have to be optimized in order to fit experimental data.
 To overcome these difficulties, approximate solutions of the SCM, valid in a limited range of parameters, have been obtained. \cite{kehr, keren}

In this paper we illustrate an effective method, alternative to both numerical quadrature and approximate solutions, to solve the SCM and calculate $G(t)$ for a generic static field distribution. 
The basic idea underlying the method is replacing Laplace integrals with discrete Fourier transforms (DFT). 
However, the naive replacement of continuous-variable integrals with discrete sums, as proposed by Weber {\it al.},  \cite{kalvius} leads to badly inaccurate results. 
In order to correctly transpose the original integral equation into a summation equation, we recalculated the Markov chains 
directly and self-consistently over a discretized time base $t_n$, in a so-called discrete-time SCM (DTSCM).
The solution of the resulting equation by DFT provides an efficient and accurate algorithm to calculate $G(t_n)$, suitable for real time fitting of experimental data.

The paper is organized as follows. The 
continuous-time SCM is recalled for reference in section \ref{sec:SCM}. Its reformulation into a DTSCM is detailed in section \ref{sec:DTSCM}. The application of the DTSCM as an effective calculation method is discussed in section \ref{sec:method}.

\section{The strong collision model }
\label{sec:SCM}

We recall here briefly the results of the original SCM applied to the muon spin evolution in a randomly distributed local field, due to Hayano {\it et al}.
\cite{hayano}  The model postulates that after a ``collision'' event, occurring with a probability $\nu$ per unit time, the local field is a random variable totally uncorrelated with the local field before the collision, and governed by the same distribution (i.e.\ collisions map the static field distribution into itself). Based on these assumptions, 
it is legitimate to treat the muon spin ensemble, described by its polarization function $G(t)$, as a single entity subject to collisions as a whole. \cite{hayano,kehr} 
$G(t)$ will then evolve as the unperturbed static-field function $G^{(0)}(t)$  with probability $\exp(-\nu t)$, corresponding to no scattering event in the $[0, t]$ time interval; or it will resume as $G^{(0)}(t_1)G^{(0)}(t-t_1)$ with probability $\lambda\exp(-\nu t)\rmd t_1$ after a collision occurred in a time interval $\rmd t_1$ around $t_1$, and so on. This leads to the following expansion in powers of $\nu$:
\begin{eqnarray}
\label{eq:cseries1}
\fl G(t)& =&  \rme^{-\nu t}G^{(0)}(t) \,+ \nonumber \\ 
\fl & & \mskip -40mu \nu\mskip -0mu \int_0^t \mskip -6mu \rmd t_1 \mskip 4mu \rme^{\nu t_1}G^{(0)}(t_1) \,\rme^{\nu(t - t_1)}G^{(0)}(t - t_1) \,+ \nonumber \\ 
\fl & &\mskip -40mu \nu^2 \mskip -6mu \int_0^t \mskip -6mu \rmd t_1 \mskip 4mu \rme^{\nu t_1}G^{(0)}(t_1) \cdot \nonumber \\  \fl & &\mskip -40mu
~\int_{t_1}^t \mskip -6mu \rmd t_2 \mskip 4mu \rme^{\nu(t_2 - t_1)}G^{(0)}(t_2 - t_1)\, \rme^{\nu(t - t_2)}G^{(0)}(t - t_2) \nonumber \\ \fl & &\mskip -40mu
\,+ ~ \dots 
\end{eqnarray}
which is rewritten into the following recursion series, upon changing the order of integration:
%
\begin{eqnarray}
\label{eq:crecursion}
\fl G(t)&  =&  \rme^{-\nu t}G^{(0)}(t) 
+ \nu \mskip -6mu \int_0^t \mskip -6mu \rmd t_1 \rme^{-\nu (t - t_1)}G^{(0)}(t - t_1)
\cdot
\nonumber \\ \fl & &\mskip -40mu
\left \{   \rme^{-\nu t_1}G^{(0)}(t_1) + 
\nu \mskip -6mu \int_0^{t_1} \mskip -6mu \rmd t_2 
\mskip 4mu \rme^{-\nu (t_1 - t_2)}G^{(0)}(t_1-t_2) \right .
\cdot \nonumber  \\ \fl &&\mskip -40mu ~ 
\left . \left [\,
\rme^{-\nu t_2}G^{(0)}(t_2) 
+ \nu \mskip -6mu \int_0^{t_2} \mskip -6mu \rmd t_3 \dots \right ] \cdot \quad \dots \quad \right \}. 
\end{eqnarray}
From the comparison with the right hand side of \eref{eq:crecursion}, it is apparent that the expression in braces equals $G(t_1)$.  
Defining 
$H^{(0)}(t) \equiv  G^{(0)}(t)\exp(-\nu t)$, 
the following Dyson-type integral equation is obtained for $G(t)$:
\numparts
\begin{equation}
\label{eq:cdyson}
\fl G(t) = H^{(0)}(t) +   \nu \mskip -6mu \int_0^t \mskip -6mu \rmd t_1 \mskip 4mu
H^{(0)}(t-t_1)\, G(t_1) 
\end{equation}
or
 \begin{equation}
\label{eq:cdysonstar}
\fl G = H^{(0)} + \nu\, H^{(0)}\mskip -6mu \ast G
\end{equation}
\endnumparts
where the convolution operator ``$\ast$'' 
is defined as the integral in the rightmost term of 
\eref{eq:cdyson}, i.e.\ as in the theory of Laplace 
transform.
Equation \neref{eq:cdysonstar} 
is solved in principle by Laplace transformation. Let ${\cal H}^{(0)}(s)\equiv {\cal L}_s\{H^{(0)}\}$, ${\cal G}^{(0)}(s)\equiv {\cal L}_s\{G^{(0)}\}$, ${\cal G}(s)\equiv {\cal L}_s\{G\}$ be the Laplace 
transforms of $H^{(0)}(t)$, $G^{(0)}(t)$ and $G(t)$, respectively; then \cite{hayano}
\begin{equation}
\label{eq:claplace}
\fl {\cal G}(s) = \frac{{\cal H}^{(0)}(s)}{1-\nu {\cal H}^{(0)}(s)} =
\frac{{\cal G}^{(0)}(s+\nu)}{1-\nu {\cal G}^{(0)}(s+\nu)}  
~ . 
\end{equation}

An analytical expression for $G(t)$ can be obtained from \eref{eq:claplace} in the case of a Lorentzian distribution of random static  
fields in zero external field, corresponding to a Lorentzian Kubo-Toyabe 
polarization function \cite{Kubo81}
\begin{equation}
\label{eq:LKT}
\fl G_{\rm L}^{(0)}(t) =\frac{1}{3}+ \frac{2}{3}(1-\Lambda t)\,\rme^{-\Lambda t}
\end{equation}
where $\Lambda/\gamma_\mu$ is the distribution half width ($\gamma_\mu$ is the muon gyromagnetic ratio). Its Laplace transform is straightforwardly calculated as 
\begin{equation}
\label{eq:LKTs}
\fl {\cal G}_{\rm L}^{(0)}(s) =\frac{1}{3s}+ \frac{2}{3(s+\Lambda)} - \frac{2}{3(s+\Lambda)^2}~.
\end{equation}
From \eref{eq:claplace} and\eref{eq:LKTs},  the $s$-domain dynamic function ${\cal G}_{\rm L}(s)$ 
is a third-order rational function with non-degenerate poles $P_k(\nu, \Lambda)$ for $\nu \neq 0$, which is decomposed into a sum of simple 
fractions of the form 
\begin{equation}
\label{eq:DLKTs}
\fl {\cal G}_{\rm L}(s)=\sum_{k=1}^3 \,\frac{C_k(\nu, \Lambda)}{s - P_k(\nu, \Lambda)}
\end{equation}
whence the time-domain 
function $G_{\rm L}(t)$ is 
the superposition of three exponentials, 
\begin{equation}
\label{eq:DLKT}
\fl G_{\rm L}(t)=\sum_{k=1}^3 C_k\, \rme^{P_kt}~.
\end{equation}

The application of the SCM  to a Lorentzian random field distribution is a rather academic exercise.
Here, the main interest of \eref{eq:DLKT} is providing an exact solution of the   continuous-time SCM to be used as a benchmark for the DTSCM developed in the next section.
In most situations of practical interest the field distribution is instead Gaussian, as in the case of the stray dipolar fields from nuclei,  which 
produce a muon 
depolarization following 
the static Kubo-Toyabe function \cite{KT1967} at low temperature,
\begin{equation}
\label{eq:KT}
\fl G_{\rm KT}^{(0)}(t) =\frac{1}{3}+ \frac{2}{3}(1-\Delta^2\,t)\,\rme^{-\frac{1}{2}\Delta^2\,t^2}
\end{equation} 
where $(\Delta/\gamma_\mu)^2$ is the second moment 
of each Cartesian component of the local 
field.  \cite{schenck}
In this context the SCM correctly describes the 
effect of the thermally activated diffusion of the muon on its relaxation function. However, an analytic expression for the \textit{dynamic} Kubo-Toyabe function $G_{\rm KT}(t)$ analogous to \eref{eq:DLKT} cannot be obtained from \eref{eq:KT} and \eref{eq:claplace}, therefore $G_{\rm KT}(t)$ has to be calculated numerically.

\section{The discrete-time strong collision model} 
\label{sec:DTSCM}

We now modify the original SCM sketched in the previous  
section \ref{sec:intro}, 
by imposing that scattering events may occur only at discrete times $t_n = n \tau$, where $n$ is an integer. 
Let $\lambda$ be the probability that 
a collision occurs over the finite time lag $\tau$, and $q=1-\lambda$ the complementary probability.
According to the above definitions, $\lambda =  1 - \exp(-\nu\tau)$,
while the probability that no collision occurs over a time $t_n$ 
equals $q^n = \exp(-n\nu\tau)$.
Following a similar argument as for the continuous-time case, the muon spin polarization $G_n\equiv G(t_n)$ will be the unperturbed  $G_n^{(0)}$ with  probability $q^n$; or it will be given by the free evolution  to $t_k$  with probability $q^{k-1}$, followed by the free evolution to $t_n$ with probability $q^{n-k}$, in the case of single collision occurred at a non-zero time $t_k \le t_n$ with probability $\lambda$, each $k$ thus contributing a term $G_{n-k}^{(0)} G_k^{(0)}$ with probability $\lambda q^{n-1}$; and so on. We are thus led to write the following equations for the Markov chain:
\begin{eqnarray}
\label{eq:dseries1}
\fl  G_0 &=& G_0^{(0)} = 1\nonumber \\ 
\fl G_1 &=& q^1 G_1^{(0)} + \lambda q^0 G_1^{(0)} \cdot q^0 G_0^{(0)} \nonumber \\ 
\fl G_2 &=& q^2 G_2^{(0)} + \lambda \left (q^0 G_1^{(0)} \cdot  q^1 G_1^{(0)} + 
q^1 G_2^{(0)} \cdot  q^0 G_0^{(0)} \right ) + \nonumber \\
\fl &&\mskip-28mu \lambda^2  q^0 G_1^{(0)} \cdot q^0 G_1^{(0)}\cdot q^0 G_0^{(0)} \nonumber \\
\fl  \dots\mskip-1mu && \nonumber \\
\fl G_n &=& q^n G_n^{(0)} + \lambda \sum_{k=0}^{n-1} q^k G_{k+1}^{(0)} \cdot q^{n-k-1} G_{n-k-1}^{(0)} + \nonumber \\
\fl &&\mskip-28mu
 \lambda^2  \sum_{k=0}^{n-1} q^k G_{k+1}^{(0)}\mskip -6mu \sum_{h=0}^{n-k-2} q^h G_{h+1}^{(0)} \cdot 
 q^{n-k-h-2}G_{n-k-h-2}^{(0)} +\nonumber \\
\fl &&\mskip-28mu\, \dots 
\end{eqnarray}
%
Defining 
$H_n^{(0)}\equiv  q^n G_n^{(0)} = G_n^{(0)}\exp(- n\nu\tau)$
as in the continuous time case, and taking into account that $H_0^{(0)}=1$,
equation \eref{eq:dseries1} is straightforwardly rewritten 
as 
\begin{eqnarray}
\label{eq:dseries2}
\fl\mskip-48mu  G_n&=& H_n^{(0)} + \frac{\lambda}{q} \sum_{k=0}^{n-1} H_{k+1}^{(0)} H_{n-k-1}^{(0)} +
\nonumber \\
\fl  &&\mskip-40mu \left (\frac{\lambda}{q}  \right )^{\!\!2}\, \sum_{k=0}^{n-1} \mskip 6mu \sum_{h=0}^{n-k-2} H_{k+1}^{(0)} H_{h+1}^{(0)} H_{n-k-h-2}^{(0)} 
+ \,\dots 
\end{eqnarray}
whence, upon factoring the outermost summation, 
 a 
recursive series is obtained, analogous of \eref{eq:crecursion}
\begin{eqnarray}
\label{eq:drecursion}
\fl \mskip 0mu G_n& =& H_n^{(0)} + \frac{\lambda}{q} \sum_{k=0}^{n-1} H_{k+1}^{(0)}
\left \{ H_{n-k-1}^{(0)} + 
\phantom{\left ( \frac{\lambda}{q}\right )^{\!\!2}\, \sum_{h=0}^{n-k-2} \mskip -40mu \sum_{j=0}^{n-k-h-3}  \mskip -260mu H_{h+1}^{(0)}H_{j+1}^{(0)} H_{n-k-h-j-3}^{(0)} }
 \right .
\nonumber \\
\fl  && \mskip-32mu 
\frac{\lambda}{q} \sum_{h=0}^{n-k-2}\mskip-6mu  H_{h+1}^{(0)} H_{n-k-h-2}^{(0)}\,+ 
\\
\fl  &&  \mskip-40mu \left . \left ( \frac{\lambda}{q}\right )^{\!\!2}\, \sum_{h=0}^{n-k-2} \mskip -16mu \sum_{j=0}^{\mskip 28mu n-k-h-3}  \mskip -16mu H_{h+1}^{(0)}H_{j+1}^{(0)} H_{n-k-h-j-3}^{(0)}  
\,+ 
\dots \right \} .\nonumber 
\end{eqnarray}
Upon recognizing that the expression within braces in \eref{eq:drecursion} equals the expansion for $G_{n-k-1}$ as of \eref{eq:dseries2}, we obtain the 
following summation equation
\numparts
\begin{eqnarray}
\label{eq:ddyson}
 \fl G_n &= &H_n^{(0)} + \frac{\lambda}{q} \sum_{k=1}^{n} H_k^{(0)} G_{n-k} =
\nonumber \\
\fl &&
H_n^{(0)} + \frac{\lambda}{q} \sum_{k=0}^{n} (H_k^{(0)} - \delta_{k,0})G_{n-k} 
\end{eqnarray}
or
\begin{equation}
\label{eq:ddysonconv}
\fl G=H^{(0)} +\frac{\lambda}{q}\, \conv\!\left(H^{(0)}\!-\delta,G \right )
\end{equation}
\endnumparts
where $\delta_n\equiv \delta_{n,0}$, and  the $\conv()$ operator is defined as 
\begin{equation}
\label{eq:dconv}
\fl \conv(A,B)_n \equiv \sum_{k=0}^n A_kB_{n-k}
\end{equation}
formally analogous to the the convolution operator ``$\ast$'' defined in section \ref{sec:SCM}. 

Equation \neref{eq:ddyson} (as well as its continuos-variable counterpart 
\eref{eq:cdyson}) exhibits a remarkable invariance by exponential 
weighting. 
Let $G$ be the solution of \eref{eq:ddyson}; then,
the same equation holds 
also for the exponentially weighted quantities $\tilde H_n^{(0)}\equiv H_n^{(0)}\exp(-\alpha n)$,
$\tilde G_n\equiv G_n\exp(-\alpha n)$:  
\begin{equation}
 \label{eq:ddysonconvwei}
\fl {\tilde G}={\tilde H}^{(0)} +\frac{\lambda}{q}\, \conv\!\left({\tilde H}^{(0)}\!-\delta,{\tilde G}\right).
\end{equation}

Despite the formal similarity 
between equations 
\eref{eq:cdysonstar} and 
\eref{eq:ddysonconv}, an exact closed expression  
in terms of $H^{(0)}$, analogous to \eref{eq:claplace}, cannot be obtained 
for the discrete time case.
Indeed, the $\conv()$ 
operator defined in \eref{eq:dconv} 
is not transformed  into a product by 
DFT (denoted hereafter as $\dft{}$). 
Rather, it is {\em circular} convolution, defined as 
 \begin{equation}
\label{eq:dcircconv}
\fl (A\ast B)_n \equiv \sum_{k=0}^{N-1} A_kB_{\modN{n-k}}
\end{equation}
where $N$ is the dimension of the discretized time base and $r=\modN{m}$ is the remainder of $m$ modulo $N$ ($0\le r < N$), 
which is transformed into a product: \cite{fft_theory} 
$\dft{A \ast B} = \dft{A}\dft{B}$.
Nonetheless,  
non-circular convolution \eref{eq:dconv} may be reduced  
to circular convolution \eref{eq:dcircconv} by doubling the space dimension and padding vectors with $N$ trailing 
zeros. Let $A$, $B$ 
be arbitrary vectors, 
$\Pad{A}$, $\Pad{B}$ 
the corresponding zero-padded vectors, 
defined as
\begin{displaymath}
\fl \Pad{A}_n= 
\begin{cases}
A_n & \text{for~} 0 \le n <  N \\
0 & \text{for~} N\le n < 2N
\end{cases}
\end{displaymath}
etc., and let $U$ be the zero-padded unit: 
$U_n=1$ for $n < N$, $U_n=0$ for $n \ge N$.   
Then
\begin{equation}
\label{eq:dconvstar}
\fl U_n\, \conv\!\left(\Pad{A},\Pad{B}\right )_n= U_n (\Pad{A}\ast \Pad{B})_n. 
\end{equation}
Henceforth we implicitly consider a doubled space dimension and zero-padded $G^{(0)}$, $H^{(0)}$ vectors, with the P superscript dropped for 
simplicity of notation. It is intended that only vector elements with indices $n<N$ are physically meaningful. Multiplying both sides of \eref{eq:ddysonconv} by $U$,
substituting \eref{eq:dconvstar} therein, and
taking into account that $U_n \Pad{A}_n=\Pad{A}_n$, we then obtain
\numparts
\begin{equation}
\label{eq:ddysonstar}
\fl G - qH^{(0)}\! = 
\lambda (H^{(0)}\! \ast G) + R \quad \left [  R_n\equiv \lambda (1 - U_n)(H^{(0)}\! \ast G)_n \right ]
\end{equation}
and a similar equation for the exponentially weighted quantity $\tilde G$
\begin{equation}
\label{eq:ddysonstarwei}
\fl \tilde G - q\tilde H^{(0)}\! = 
\lambda (\tilde H^{(0)}\! \ast \tilde G) + R' \quad \!\left [  R_n'\equiv \lambda (1 - U_n)(\tilde H^{(0)} \!\ast \tilde G)_n \right ]
\end{equation}
\endnumparts

Due to the $R$ term on the right hand side of \eref{eq:ddysonstar}, a closed exact expression for $\dft{G}$ as a function of $\dft{H^{(0)}}$ cannot be obtained yet. 
In the analogous \eref{eq:ddysonstarwei} for $\tilde G$, however, the ``error'' $R'$ can be made arbitrarily small by an arbitrarily large weighting exponent $\alpha$. This suggests that the approximate solution $\tilde G'$ of \eref{eq:ddysonstarwei} obtained by dropping $R'$ 
is asymptotically exact. 
A more rigorous proof that $G_n'\equiv  \tilde G_n'\exp(\alpha n)$ actually tends to $G_n$ for $\alpha \to \infty$ is outlined in \ref{sec:proof}.
The approximate solution
$\dft{\tilde G_n'}$ 
is 
then
straightforwardly written in the following closed form
\begin{equation}
\label{eq:dfourier}
\fl  \dft{\tilde G'} =
\frac{q\dft{\tilde H^{(0)}}}{1 -\lambda\dft{\tilde H^{(0)}}} \approx \dft{\tilde G} . 
 \end{equation}

Equation \neref{eq:dfourier}, which is the discrete-time analogous of \eref{eq:claplace}, is the main result of this paper. Its practical implementation in an effective calculation algorithm for $G(t)$ is discussed in the next section.

\section{Application of the DTSCM method }
\label{sec:method} 

Summarizing the above results, 
the muon longitudinal 
polarization function $G_n=G(t_n)$,  $0 \le n < N$ in the DTSCM approximation  is calculated as  

\begin{eqnarray}
\label{eq:algorithm}
\fl\mskip -0mu  G_n &\approx &G'_n = \,\rme^{\alpha n} \times 
\nonumber \\
  &&\mskip -28mu 
\idft{\frac{\rme^{-\nu\tau}\dft{\rme^{-(\nu\tau + \alpha)m}\, G_m^{(0)}}}
{1 -(1-\rme^{-\nu\tau})\dft{ \rme^{-(\nu\tau + \alpha)m}\,G_m^{(0)}}}}_n  \mskip 40mu ~
 \end{eqnarray}
with $G_m^{(0)}$,  $0 \le m < 2N$ being the zero-padded static relaxation function defined such that $G_m^{(0)} = 0$ for $m\ge N$, and $\idft{}$ the inverse DFT. The weighting coefficient  $\alpha$ 
is a large-enough positive quantity, whose practical choice is discussed in the following. In high-level mathematics-oriented computer programming languages such as Matlab or Octave, which provide the fast Fourier transform (FFT) 
built-in or in a standard library,   
\eref{eq:algorithm} is implemented by just a few lines of code \footnote{Sample routines running under Matlab and Octave are made available online in the Supplemental Materials accompanying this paper.}.

The accuracy of  $G'_n$ 
as an approximation for the exact solution $G_n$ of \eref{eq:ddysonconv} depends critically on the proper tuning of the exponential weighting.
While in principle $G'_n - G_n$ tends to zeros as $\exp(-2N\alpha)$ for 
$\alpha\to\infty$ (see Appendix \ref{sec:proof}), an exceedingly large value of the weighting coefficient $\alpha$ leads in practice to numerical overflow. On the other hand, a too small $\alpha$ brings about an error which may become very large in some particular case. In order to 
guide a convenient choice of 
$\alpha$ in \eref{eq:algorithm}, 
we compared $G'$ with the exact solution $G$ of the DTSCM corresponding to the static Kubo-Toyabe function \eref{eq:KT} for several values of $\alpha$, $\Delta$, and $\nu$. The exact $G\equiv G_{\rm KT}$ can be calculated in the general case from the expansion \eref{eq:dseries1} (or equivalently \eref{eq:dconvserieswei}) in powers of $\lambda/q$. We stress however that the summation of the series \eref{eq:dconvserieswei} constitutes a quite inefficient algorithm, as its numerical convergence requires up to nearly as many terms as $N$ for large $\nu$ values. 
In the limit case $\Delta=0$ (i.e.\ $ G_n^{(0)} = 1$) it is apparent from \eref{eq:dseries1} that identically $G_n = 1$, in agreement with the observation that dynamics cannot alter the muon polarization in the absence of an internal field.
The numerical tests were performed in standard double precision IEEE-754 floating point arithmetics \cite{ieee754} on an Intel-based personal computer running Matlab. 
The mean value and standard deviation of $G'_n - G_n$, $0 \le n <N=8192$, are listed in table \ref{tab:alpha_tunining}
for a few representative parameter values. The best accuracy, approximately $10^{-9}$, is obtained for $\alpha$ in the order of $10/N$, while for $\alpha>20/N$ the calculation of 
$G'_n$ is increasingly afflicted by floating point truncation errors, up to a numerical divergence at  $\alpha \ge 40/N$.
This trend was reproduced in all our simulations. Based on these figures, we 
chose  $\alpha = 10/N$ as a convenient setting in all the following examples.

\begin{table}
\caption{\label{tab:alpha_tunining}Average value and standard deviation of $G'_n -G_n$, where $G'_n$ and $G_n$ are, respectively,  the approximate \neref{eq:algorithm} and exact \neref{eq:dseries1} dynamic functions, calculated on $N=8192$ sampling points
from the static Kubo-Toyabe function \eref{eq:KT} in the DTSCM, for selected $\alpha$, $\nu$, and $\Delta$ values.} 
\begin{ruledtabular}
\begin{tabular}{@{}*{5}{l}}
$\alpha N$ & $\nu\tau$ & $\Delta\,\tau$ & $\left . \0\overline{G'-G} \right .$& 
$\sigma_{G'-G}$ \cr 
\hline
\0$0$  & $0.001$ & 0     & \m$2.2\times 10^{2}$  &  $1.4\times 10^{-1} $ \cr
\0$0$  & $0.01$  & 0.001 & \m$2.0\times 10^{-2}$ &  $9.4\times 10^{-3} $ \cr
\0$0$  & $0.1$   & 0     & $-8.7\times 10^{11}$  &  $1.4\times 10^{-1} $ \cr
\0$2$  & $0.001$ & 0     & \m$1.9\times 10^{-2}$ &  $1.2\times 10^{-5} $ \cr
\0$2$  & $0.01$  & 0.001 & \m$3.6\times 10^{-4}$ &  $1.7\times 10^{-7} $ \cr
\0$2$  & $0.1$   & 0     & \m$1.9\times 10^{-2}$ &  $1.4\times 10^{-13} $ \cr
\0$5$  & $0.001$ & 0     & \m$4.5\times 10^{-5}$ &  $3.0\times 10^{-8} $ \cr
\0$5$  & $0.01$  & 0.001 & \m$8.9\times 10^{-7}$ &  $4.1\times 10^{-7} $ \cr
\0$5$  & $0.1$   & 0     & \m$4.5\times 10^{-5}$ &  $2.0\times 10^{-13} $ \cr
$10$   & $0.001$ & 0     & \m$2.1\times 10^{-9}$ &  $1.4\times 10^{-12} $ \cr
$10$   & $0.01$  & 0.001 & \m$4.0\times 10^{-11}$&  $1.9\times 10^{-11} $ \cr
$10$   & $0.1$   & 0     & \m$2.1\times 10^{-9}$ &  $1.2\times 10^{-11} $ \cr
$20$   & $0.001$ & 0     & $-4.1\times 10^{-10}$ &  $3.1\times 10^{-9} $ \cr
$20$   & $0.01$  & 0.001 & \m$3.9\times 10^{-10}$&  $2.6\times 10^{-9} $ \cr
$20$   & $0.1$   & 0     & $-1.5\times 10^{-9}$  &  $7.2\times 10^{-9} $ \cr
$50$   & $0.001$ & 0     & \m$7.1\times 10^{2}$  &  $1.1\times 10^{4} $ \cr
$50$   & $0.01$  & 0.001 & \m$1.4\times 10^{3}$  &  $2.6\times 10^{4} $ \cr
$50$   & $0.1$   & 0     & $-5.9\times 10^{3}$   &  $3.3\times 10^{4} $ \cr
\end{tabular}
\end{ruledtabular}
\end{table}

\begin{figure}
\includegraphics[width=\columnwidth]{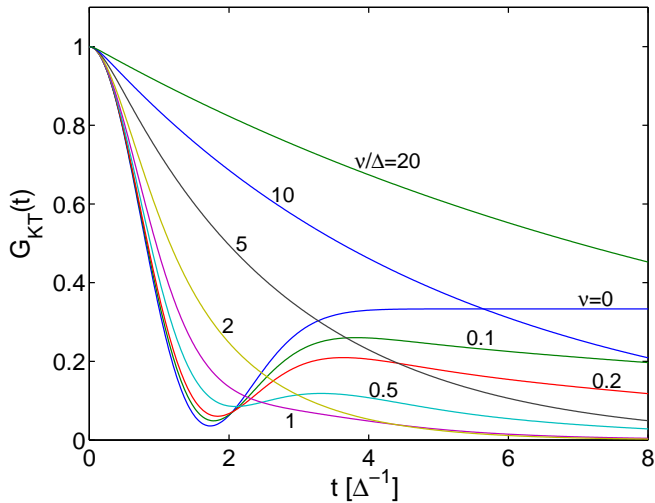}
\caption{\label{fig:hayano}
Dynamic Kubo-Toyabe function calculated by \eref{eq:algorithm} on $N=8192$ sampling points for several scattering frequencies $\nu$. }
\end{figure}

In order to benchmark the computational efficiency of the DTSCM-based method, 
we calculated the dynamical Kubo-Toyabe function for the same 
$\nu/\Delta$ values as in figure 3(a) of Hayano {\it et al.}\cite{hayano}\
on several personal computers. 
The $G_{\rm KT}(t)$ curves,  calculated 
on an array of $N=8192$ sampling points (the typical histograms length e.g.\ of the datasets from the Paul Scherrer Institute muon facility), are plotted in Fig.~\ref{fig:hayano}. The calculation in Matlab took a $0.4$~s overall CPU time on an AMD Athlon processor at 750~MHz dating back to year 2000, and approximately one tenth on a recent PC (Pentium G2030 CPU at 3.0~GHz). A fit of real $\mu$SR data, requiring 
typically several hundreds function calls, can be  therefore performed by means of \eref{eq:algorithm} virtually in real time even on a very low-end computer.

\begin{figure}
\includegraphics[width=\columnwidth]{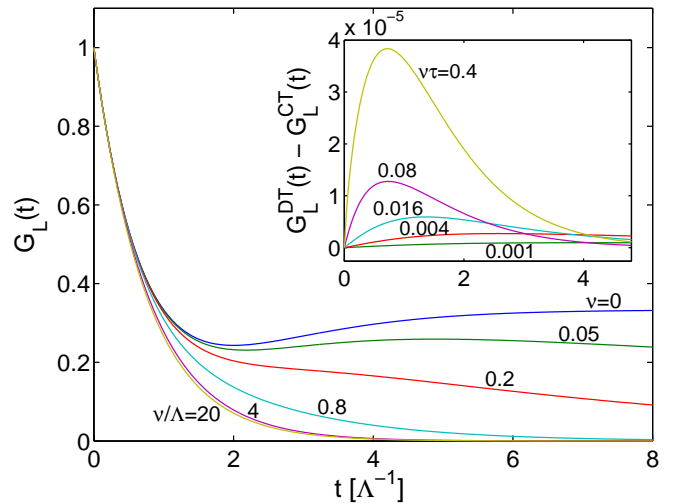}
\caption{\label{fig:LKT_DTvsCT}
Lorentzian field distribution. Main panel: dynamic polarization function $G_{\rm L}(t)$ for several scattering frequencies $\nu$.  Inset: difference between $G_{\rm L}(t)$ calculated by the DTSCM method \eref{eq:algorithm} over $N=512$ sampling points, and the exact continuous-time solution \eref{eq:DLKT}, for the various $\nu$ values.}
\end{figure}

The accuracy of the discrete-time approximation with a reasonable sampling interval, 
possibly identical to the native  experimental resolution in the time-differential data, is another issue of our method. To this end, we tested the DTSCM solution against analytical or approximate solutions of the continuous-time SCM in two cases. The first benchmark is provided by the SCM in the presence of a Lorentzian field distribution, whose exact solution is given by \eref{eq:DLKT}. The polarization function $G_{\rm L}(T)$ is plotted in Fig.~\ref{fig:LKT_DTvsCT} for several values of the scattering frequency $\nu$. The discrete-time ($N=512$) and continuous-time solutions $G_{\rm L}^{\rm DT}$, $G_{\rm L}^{\rm CT}$ are practically overlapped and undistinguished in the plot. In spite of the rather coarse time sampling, their difference  $G_{\rm L}^{\rm DT}(t)- G_{\rm L}^{\rm CT}(t)$ (figure inset) is 
negligible for practical purposes even at comparatively high $\nu$.
For reference, 
the experimental relative uncertainty on the muon polarization in very-high-statistics measurements is seldom smaller than a few permil. 

\begin{figure}
\includegraphics[width=\columnwidth]{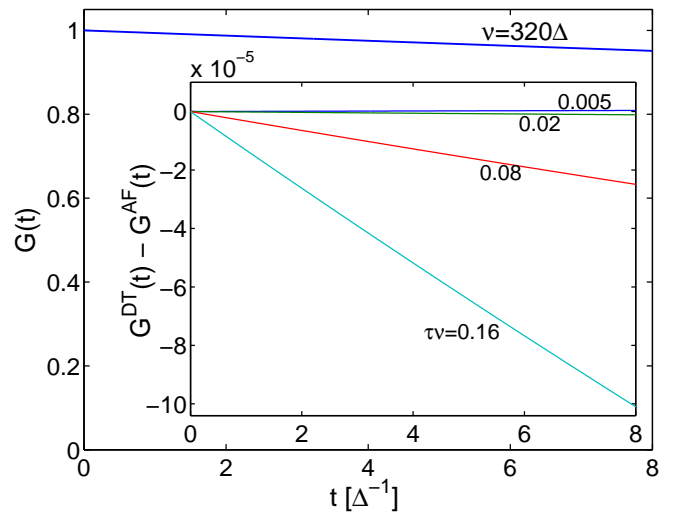}
\caption{\label{fig:narrowing_DTvsAK}
Gaussian field distribution in the extreme narrowing limit ($\nu=320\Delta$). Main panel: dynamic polarization function $G(t)$.
Inset: difference between $G(t)$ calculated by the DTSCM-based (DT) method \eref{eq:algorithm} and the Abragam formula (AF) asymptotic solution \eref{eq:Keren}, for several sampling intervals $\tau$.}
\end{figure}

Another well-known case is the Gaussian field distribution in the extreme narrowing limit $\nu \gg\Delta $. Its relaxation function is approximated  by the so-called Abragam formula \cite{abragam,keren, kehr}
\begin{equation}
 \label{eq:Keren}
 \fl G^{\rm AF}(t)= \exp(-2 \Delta^2\nu^{-2})  \left ( \exp(-\nu t)- 1 + \nu t \right)
\end{equation}
which is asymptotically exact for $\nu/\Delta \to \infty$. The accuracy of \eref{eq:algorithm} in reproducing \eref{eq:Keren} is exemplified in Fig.~\ref{fig:narrowing_DTvsAK}, showing simulations  obtained by the DTSCM at
 $\nu=320\Delta$ and
different time resolutions. Here again, the difference  $G^{\rm DT}(t)- G^{\rm AF}$ between the discrete-time and the continuous-time solution given by the Abragam approximate formula is negligible even with a relatively coarse-grained sampling,  corresponding to a cumulative scattering probability over a time bin $1-\exp(-\nu\tau)$ on the order of ten percent.

\section{Conclusions}
\label{sec:discussion}
In conclusion, we have demonstrated an accurate and efficient numerical method to calculate the dynamical Kubo-Toyabe function describing the longitudinal muon polarization function $G(t)$ vs.\ time in the presence of muon diffusion as well as, in principle, the solution of the SCM for an arbitrary distribution of static internal fields. 
The error on $G(t)$ produced by time discretization is found to be much smaller than 
the experimental uncertainty even with very coarse time resolutions, and data oversampling is not needed.
If implemented by means of the FFT algorithm, the method requires negligible CPU resources, which makes it suitable to fit experimental data in real time even on a slow computer.

\section*{ACKNOWLEDGEMENT}
The authors are indebted with G.\ Guidi and C.\ Bucci for helpful and stimulating discussion.

\appendix

\section{Proof that $ G'$ tends to $G$ for $\alpha\to\infty$}
\label{sec:proof}
We sketch here the proof that  $G'$, defined by  \eref{eq:algorithm}, 
is an asymptotically exact solution of \eref{eq:ddysonconv} for $\alpha\to\infty$. It is easily shown 
that the exponentially weighted function $\tilde G'_n$ ($0\le n< 2N$) defined by  \eref{eq:dfourier} obeys the equation
\begin{equation}
 \label{eq:ddysonstareqn}
\fl {\tilde G'}={\tilde H}^{(0)} +\frac{\lambda}{q}\, ({\tilde H}^{(0)}-\delta)\ast {\tilde G'}
\end{equation}
formally identical to \eref{eq:ddysonconvwei} but for the replacement of non-circular with circular convolution. Upon defining ${\tilde K}^{(0)}\equiv {\tilde H}^{(0)}\!-\delta$, the following series  expansion is straightforwardly obtained from \eref{eq:ddysonstareqn}:
 \begin{eqnarray}
 \label{eq:dstarserieswei}
\fl {\tilde G'}&= &{\tilde H}^{(0)} +\frac{\lambda}{q}\, {\tilde K}^{(0)}\!\ast {\tilde H}^{(0)} + 
\frac{\lambda^2}{q^2}\, {\tilde K}^{(0)}\!\ast {\tilde K}^{(0)}\!\ast {\tilde H}^{(0)} + \nonumber \\
\fl &  &\mskip -32mu
\frac{\lambda^3}{q^3}\, {\tilde K}^{(0)}\!\ast {\tilde K}^{(0)}\!\ast {\tilde K}^{(0)}\!\ast {\tilde H}^{(0)}
+  \dots \nonumber \\
\fl &  
\equiv & \sum_{m=0}^\infty {\tilde {\cal A}}^{(m)}
\end{eqnarray}
to be compared with the analogous expansion for the exact solution ${\tilde G}_n$  ($0\le n < N$) drawn from \eref{eq:ddysonconvwei},
\begin{eqnarray}
 \label{eq:dconvserieswei}
\fl {\tilde G}&= &{\tilde H}^{(0)} +\frac{\lambda}{q}\, \conv\!\left({\tilde K}^{(0)}, {\tilde H}^{(0)}\right) + \nonumber \\\fl &&
\mskip -32mu \frac{\lambda^2}{q^2}\, \conv\!\left({\tilde K}^{(0)}, \,\conv\!\left({\tilde K}^{(0)}, {\tilde H}^{(0)}\right) \right)+  \nonumber \\
\fl &  &
\mskip -32mu \frac{\lambda^3}{q^3}\, \conv\!\left({\tilde K}^{(0)}, \,
\conv\!\left({\tilde K}^{(0)}, \,\conv\!\left({\tilde K}^{(0)}, {\tilde H}^{(0)}\right) \right) \right) 
+  \dots \nonumber \\
\fl &\equiv &\sum_{m=0}^\infty {\tilde {\cal C}}^{(m)}
\end{eqnarray}
 which is actually a finite summation, as 
${\tilde {\cal C}}^{(m)}$
 vanish identically for $m>N$ (see \eref{eq:dseries1}). 

The series \eref{eq:dstarserieswei} is  absolutely convergent for any positive $\alpha$.  
Its $m$-th term ${\tilde {\cal A}}^{(m)}$ 
clearly obeys the recursion relation 
\begin{equation}
\label{eq:gmrecursion}
\fl {\tilde {\cal A}}^{(m+1)} = \frac{\lambda}{q}\,{\tilde K}^{(0)}\!\ast {\tilde {\cal A}}^{(m)}.
\end{equation}
Taking into account that 
$\abs{ {\tilde K_n}^{(0)}}\le\exp(-(\alpha+\nu\tau)n)$, from \eref{eq:gmrecursion} we can set the following  upper bound:
\begin{eqnarray}
\label{eq:gmconverge}
\fl \abs{{\tilde {\cal A}_n}^{(m+1)}}& \le& \frac{\lambda}{q}\sum_{k=1}^{N-1} 
\rme^{-(\alpha+\nu\tau)n}\max{{\tilde {\cal A}}^{(m)}} \nonumber \\
\fl &&\mskip -76mu = \rme^{-\alpha}\left( 1 - \rme^{-(\alpha+\nu\tau)(N-1)} \right )
\frac{1-\rme^{-\nu\tau}}{1- \rme^{-(\alpha+\nu\tau )}} \max{{\tilde {\cal A}}^{(m)}} \nonumber \\
\fl & & \mskip -76mu <  \rme^{-\alpha}\max{{\tilde {\cal A}}^{(m)}}
\end{eqnarray}
whence, by induction, 
$ \abs{{\tilde {\cal A}_n}^{(m)}} < A\exp({-\alpha m})$
with $A$ being a suitable positive constant. This ensures the absolute convergence of the series.

We now evaluate the difference 
${\tilde  E}_n \equiv  {\tilde G}'_n - {\tilde G}_n$ ($0\le n < N$) term by term from \eref{eq:dstarserieswei} and \eref{eq:dconvserieswei}:
  \begin{equation}
\label{eq:errweiseries}
\fl {\tilde  E}_n = \sum_{m=2}^\infty {\tilde {\cal E}}_n^{(m)}=\sum_{m=2}^\infty  {\tilde {\cal A}}_n^{(m)} -  {\tilde {\cal C}}_n^{(m)}
\end{equation}
where the first non-zero term in the sum is $m=2$, since $ {\tilde {\cal A}}_n^{(1)} =  {\tilde {\cal C}}_n^{(1)}$ owing to \eref{eq:dconvstar}.
The $m$-th term ${\tilde {\cal A}}_n^{(m)}$  in \eref{eq:dstarserieswei} is expressed as 
\begin{equation}
\label{eq:calAterm}
\fl {\tilde  {\cal A}}_n^{(m)} = \frac{\lambda^m}{q^m}\underbrace{\sum_{k=0}^{N-1} 
\dots\sum_{s=0}^{N-1}}_{m\ {\rm times}} 
{\tilde K}^{(0)}_k
\dots {\tilde K}^{(0)}_s {\tilde H}_{\moddN{n-k\, \dots -s}}^{(0)} 
\end{equation}
where the summation indices are upper-limited to $N-1$ since 
${\tilde K}^{(0)}$ is a zero-padded vector. On the other hand,
${\tilde {\cal C}}_n^{(m)}$ from \eref{eq:dconvserieswei} is calculated as 
\begin{equation}
\label{eq:calCterm}
\fl {\tilde  {\cal C}}_n^{(m)} = \frac{\lambda^m}{q^m}\underbrace{\sum_{k=0}^{N-1} 
\dots\sum_{s=0}^{N-1}}_{m\ {\rm times}} 
{\tilde K}^{(0)}_k
\dots {\tilde K}^{(0)}_s {\tilde H}_{n-k\, \dots -s}^{(0)}  
\end{equation}
where it is intended that ${\tilde H}_j^{(0)}=0$ for $j < 0$.
Equations \neref{eq:calAterm} and \neref{eq:calCterm} may also be written as 
\begin{eqnarray}
\label{eq:calAtermdelta}
\fl {\tilde  {\cal A}}_n^{(m)}\! &= &\frac{\lambda^m}{q^m}\overbrace{\sum_{k=0}^{N-1} 
\dots\sum_{s=0}^{N-1}}^{m\ {\rm times}} 
\sum_{t=0}^{N-1}
{\tilde K}^{(0)}_k
\dots {\tilde K}^{(0)}_s {\tilde H}_{t}^{(0)}\times
 \\ \fl && \mskip -40mu 
\left(\delta_{n-k\, \dots -s, t} +
 \delta_{2N+n-k\, \dots -s, t} +  \delta_{4N+n-k\, \dots -s, t} + \dots \right)\nonumber
\end{eqnarray}
and 
\begin{equation}
\label{eq:calCtermdelta}
\fl {\tilde  {\cal C}}_n^{(m)} = \frac{\lambda^m}{q^m}\underbrace{\sum_{k=0}^{N-1} 
\dots\sum_{s=0}^{N-1}}_{m\ {\rm times}} 
\sum_{t=0}^{N-1}
{\tilde K}^{(0)}_k
\dots {\tilde K}^{(0)}_s {\tilde H}_{t}^{(0)}\delta_{n-k\, \dots -s, t}  
\end{equation}
respectively, whence ${\tilde  {\cal E}}_n^{(m)}$ is expressed as
\begin{widetext}
\begin{equation}
\label{eq:ERRtermdelta}
\fl {\tilde  {\cal E}}_n^{(m)} = \frac{\lambda^m}{q^m}\overbrace{\sum_{k=0}^{N-1} 
\dots\sum_{s=0}^{N-1}}^{m\ {\rm times}} 
\sum_{t=0}^{N-1}
{\tilde K}^{(0)}_k
\dots {\tilde K}^{(0)}_s {\tilde H}_{t}^{(0)}
\left (\delta_{2N+n-k\, \dots -s, t} + \delta_{4N+n-k\, \dots -s, t} + \dots \right).%
\end{equation}
The latter expression is subject to the following bound:
\begin{eqnarray}
\label{eq:ERRtermbound}
\fl \abs{{\tilde  {\cal E}}_n^{(m)}} &\le &\frac{\lambda^m}{q^m}\overbrace{\sum_{k=0}^{N-1} 
\dots\sum_{s=0}^{N-1}}^{m\ {\rm times}} 
\sum_{t=0}^{N-1}
\rme^{-k(\alpha+\nu\tau)}\dots \rme^{-s(\alpha+\nu\tau)} \rme^{-t(\alpha+\nu\tau)} 
\left (\delta_{2N+n-k\, \dots -s, t} + \delta_{4N+n-k\, \dots -s, t} + \dots 
\right ) \nonumber \\
\fl ~ & = & \frac{\lambda^m}{q^m}\,\rme^{-n(\alpha+\nu\tau)}\!
\sum_{k=0}^{N-1} 
\dots\sum_{s=0}^{N-1}\sum_{t=0}^{N-1} 
\left (\delta_{2N+n-k\, \dots -s, t}\,\rme^{-2N(\alpha+\nu\tau)} 
+ 
\delta_{4N+n-k\, \dots -s, t}\,\rme^{-4N(\alpha+\nu\tau)} + \dots 
\right )  \nonumber \\
\fl & = &\frac{\lambda^m}{q^m}\,\rme^{-(n+2N)(\alpha+\nu\tau)}\!  \sum_{h=1}^{m/2}
{\cal N}(m, n + 2hN)\, \rme^{-2(h-1)N(\alpha+\nu\tau)} 
\end{eqnarray}
\end{widetext}
where ${\cal N}(m,n)$ denotes the number of combinations whereby the sum of $m+1$ integer addends each in the range $[0, N-1]$ may yield $n$. It follows that the  unweighted difference $E_n = G'_n - G_n = {\tilde  E}_n\exp(\alpha n)$ is bounded as 
\begin{eqnarray}
\label{eq:totERRbound}
\fl \abs{ E_n} &\le & 
\rme^{-2N\alpha}\, \rme^{-(n+2N)\nu\tau}
\sum_{m=2}^\infty
\frac{\lambda^m}{q^m} \times \phantom{\mskip 120mu ~}\nonumber \\
\fl &  &\left [ \sum_{h=1}^{m/2}
{\cal N}(m, n + 2hN)\, \rme^{-2(h-1)N(\alpha+\nu\tau)}  \right]
\end{eqnarray}
where the expression on the right hand side 
tends to zero for $\alpha\to\infty$, since the series (which is convergent
in view of \eref{eq:gmconverge}) is a  decreasing function of $\alpha$, while its prefactor vanishes. 
Therefore $\lim_{\alpha\to \infty}E_n=0$.


\bibliography{DiscreteTimeSCM_arxiv}%
\end{document}